\documentclass[%
 reprint,
superscriptaddress,
preprintnumbers,
nofootinbib,
 amsmath,amssymb,
 aps,
]{revtex4-2}

\usepackage{graphicx}
\usepackage{dcolumn}
\usepackage{bm}
\usepackage{hyperref}
\usepackage{braket}
\usepackage{xcolor}
\usepackage{here}

\begin{document}

\preprint{KUNS-3073, YITP-25-145}

\title{Searching for dark photon dark matter from terrestrial magnetic fields}

\author{Kimihiro Nomura}
\email{k.nomura@tap.scphys.kyoto-u.ac.jp}
\affiliation{
Department of Physics, Kyoto University, Kyoto 606-8502, Japan
}%
\author{Atsushi Nishizawa}
\affiliation{Physics Program, Graduate School of Advanced Science and Engineering, Hiroshima University, Higashi-Hiroshima, Hiroshima 739-8526, Japan}
\affiliation{Astrophysical Science Center, Hiroshima University, Higashi-Hiroshima, Hiroshima 739-8526, Japan}
\author{Atsushi Taruya}
\affiliation{Center for Gravitational Physics and Quantum Information, Yukawa Institute for Theoretical Physics, Kyoto University, Kyoto 606-8502, Japan}
\affiliation{Kavli Institute for the Physics and Mathematics of the Universe, Todai Institutes for Advanced Study, The University of Tokyo, (Kavli IPMU, WPI), Kashiwa, Chiba 277-8583, Japan}
\affiliation{Korea Institute for Advanced Study, 85 Hoegiro, Dongdaemun-gu, Seoul 02455, Republic of Korea}
\author{Yoshiaki Himemoto}
\affiliation{Department of Liberal Arts and Basic Sciences, College of Industrial Technology, Nihon University, Narashino, Chiba 275-8576, Japan}

\date{\today}

\begin{abstract}
We present a novel search for dark photon dark matter (DM) using terrestrial magnetic field measurements at frequencies below 100 Hz. 
Coherently oscillating dark photon DM can induce a monochromatic magnetic field via kinetic mixing with ordinary photons.
Notably, for dark photon masses $m_{A'}$ around $3 \times 10^{-14} \, \text{eV}$, the signal can be resonantly amplified within a cavity formed by the Earth's surface and the ionosphere.
We compute the expected signal incorporating the effect of atmospheric conductivity, and derive new upper limits on the kinetic mixing parameter $\varepsilon$ from long-term geomagnetic data. 
These limits improve upon previous ground-based constraints in the mass range of $1 \times 10^{-15} \, \text{eV} \lesssim m_{A'} \lesssim 2 \times 10^{-13} \, \text{eV}$.
\end{abstract}

\maketitle


\section{Introduction}
The presence of an unknown, invisible component of matter, commonly referred to as dark matter (DM), is supported by a wide range of independent astronomical observations.
Uncovering the nature of DM remains one of the most pressing open problems in particle physics and cosmology.
A well-motivated class of DM candidates consists of bosonic fields with ultralight masses, including both scalar and vector fields. 
In particular, fields with masses in the range of $10^{-23}$ to $1~\text{eV}$ that undergo coherent oscillations can behave as cold DM on cosmological scales \cite{Hu:2000ke, Hui:2016ltb, Nelson:2011sf, Arias:2012az, Ferreira:2020fam}.

A promising approach to detect these ultralight DM candidates is to exploit their couplings to photons.
For instance, an axion-like scalar field $a$ is expected to interact with photons via the Lagrangian $\mathcal{L}_{\text{int}} \propto g_{\text{a}\gamma} a F_{\mu\nu} \tilde{F}^{\mu\nu}$, where $F_{\mu\nu}$ is the electromagnetic tensor, $\tilde{F}_{\mu\nu}$ is its dual, and $g_{\text{a}\gamma}$ is the coupling constant \cite{DiLuzio:2020wdo, Svrcek:2006yi, Jaeckel:2010ni}.
Another compelling DM candidate is a massive vector field---referred to as the \textit{dark photon}---which can arise from an additional $U(1)$ gauge symmetry beyond the Standard Model \cite{Holdom:1985ag, fabbrichesi2021physics, Jaeckel:2010ni}, and constitutes the primary focus of this work.
The dark photon can kinetically mix with the ordinary photon as $\mathcal{L}_{\text{int}} \propto \varepsilon F'_{\mu\nu} {F}^{\mu\nu}$, where $F'_{\mu\nu}$ is the dark photon field strength tensor, and $\varepsilon$ is the kinetic mixing parameter. 
These couplings enable the conversion of such DM candidates into ordinary photons, potentially generating detectable electromagnetic signals.
Although these signals are expected to be faint, they are actively sought by highly sensitive terrestrial experiments and astrophysical observations.
For example, laboratory-based haloscope experiments are sensitive to masses in the range of $10^{-7}$--$10^{-4}\, \text{eV}$, and have placed upper limits on the dark photon kinetic mixing parameter as $\varepsilon \lesssim 10^{-15}$ in the most constraining cases
(see \cite{AxionLimits, Caputo:2021eaa} for summaries).

In this paper, we present a novel search for dark photon DM with masses $m_{A'} \lesssim 2 \times 10^{-13} \, \text{eV}$, utilizing terrestrial magnetic field measurements at frequencies $\lesssim 100 \,\text{Hz}$. 
Coherently oscillating dark photon DM surrounding the Earth can induce a nearly monochromatic electromagnetic field via the kinetic mixing $\varepsilon$, potentially producing observable signals in geomagnetic data.
While axion DM can similarly generate monochromatic electromagnetic signals, a crucial difference is that axion signals rely on the presence of a background geomagnetic field \cite{Arza:2021ekq, Taruya:2025zql, Taruya:2025bhe}, whereas dark photon signals do not.
This detection concept was first proposed in \cite{Fedderke:2021aqo}, and searches using SuperMAG data have been conducted at frequencies $f < \text{min}^{-1}$ \cite{Fedderke:2021rrm} and $f < \text{sec}^{-1}$ \cite{Friel:2024shg}, placing upper limits on $\varepsilon$ for $m_{A'} \lesssim 4 \times 10^{-15}~\text{eV}$.
At higher frequencies around $10~\text{Hz}$, the electromagnetic wavelength becomes comparable to the Earth's size.
Interestingly, this can lead to resonant amplification of the signal within an effective cavity formed by the Earth's surface and the ionosphere.
However, in this regime, the finite conductivity of the atmosphere significantly influences the signal's peak amplitude near resonance.
In this work, we successfully extend the signal prediction to these higher frequencies by incorporating atmospheric conductivity effects, following a formalism previously developed for axion DM searches by some of the present authors \cite{Taruya:2025zql, Nishizawa:2025xka, Taruya:2025bhe}.

\begin{figure}[tb]
    \centering
    \includegraphics[width=\columnwidth]{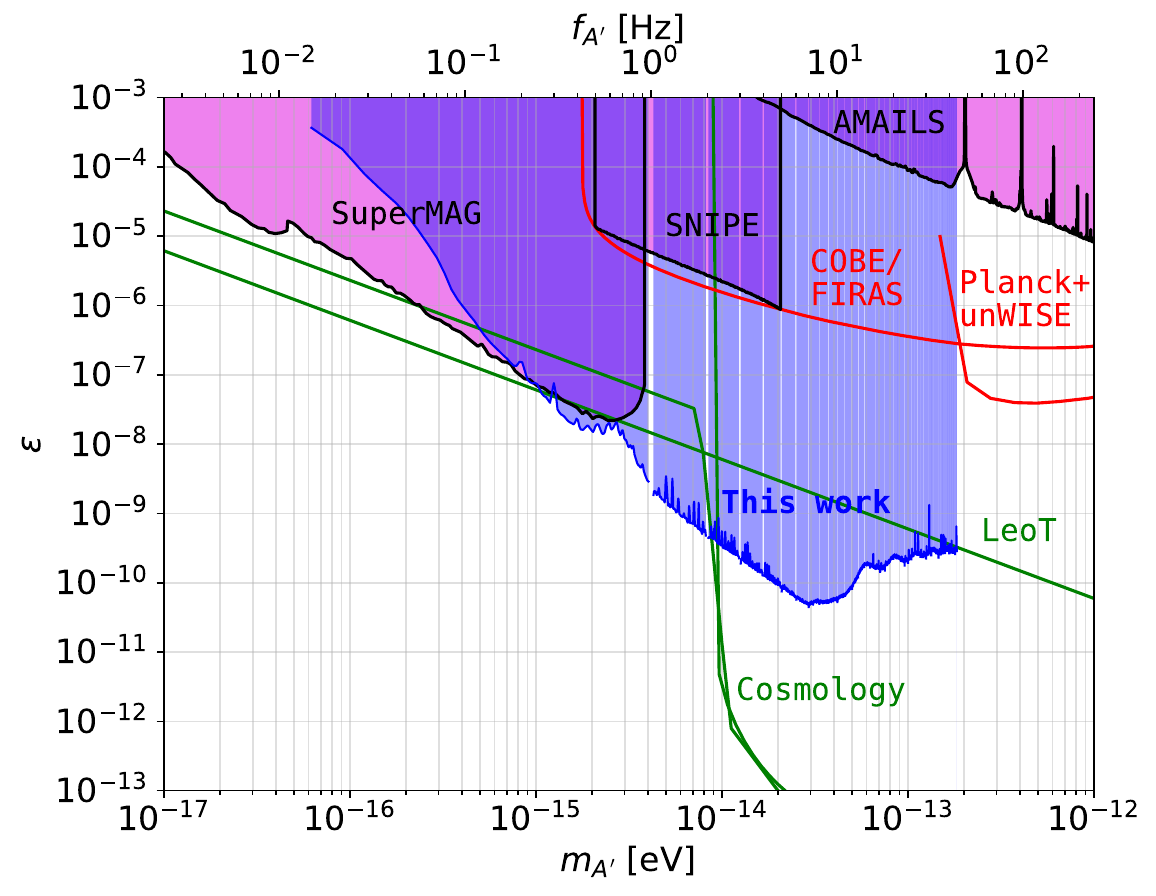}
    \caption{The blue region shows the 95\% C.L.~constraints on the kinetic mixing parameter $\varepsilon$ between dark photons and ordinary photons obtained in this work using geomagnetic data from Eskdalemuir Observatory. Here, we assume the dark photon DM energy density $\rho_{A'} = 0.3~\text{GeV}/\text{cm}^3$.
    We also show constraints from  previous studies: the magenta regions are results from ground-based direct searches (SuperMAG \cite{Fedderke:2021rrm, Friel:2024shg}, SNIPE \cite{Sulai:2023zqw}, and AMAILS \cite{Jiang:2023jhl}); the green and red curves are indirect astrophysical or cosmological bounds for dark photon DM (heating of the Leo T dwarf galaxy \cite{Wadekar:2019mpc} and other cosmological probes \cite{Caputo:2020bdy, Witte:2020rvb}) and those for non-DM dark photon (COBE/FIRAS \cite{Caputo:2020bdy, Arsenadze:2024ywr} and Planck+unWISE \cite{McCarthy:2024ozh}), respectively.
    The data are taken from \cite{AxionLimits}.}
    \label{fig:darkphoton_constraint}
\end{figure}

In summary, employing the novel signal prediction and a long-term geomagnetic dataset, we explore dark photon DM in the mass range up to $2 \times 10^{-13}\,\text{eV}$.
No significant signals are found across most frequencies, and these results are interpreted as constraints on the mixing parameter $\varepsilon$.
As shown in Fig.~\ref{fig:darkphoton_constraint}, assuming the fiducial DM energy density, we obtain stringent upper limits on $\varepsilon$ that improve upon existing ground-based direct searches in the mass range of $1 \times 10^{-15} \, \text{eV} \lesssim m_{A'} \lesssim 2 \times 10^{-13} \, \text{eV}$.

\section{Dark photon DM}

We consider a massive dark photon DM denoted by $A'_\mu$, which kinetically mixes with the ordinary (Standard Model) photon $A_\mu$. 
The action is given by 
\begin{align}
    S[A'_\mu, A_\mu] 
    &= \int d^4 x \, \bigg(
        - \frac{1}{4} F'_{\mu \nu} F^{\prime \mu \nu} - \frac{1}{4} F_{\mu \nu} F^{\mu \nu}
        \notag \\
    &\qquad
        - \frac{\varepsilon}{2} F'_{\mu \nu} F^{\mu \nu} - \frac{1}{2} m_{A'}^2 A'_\mu A^{\prime \mu}
    \bigg) \,, 
    \label{eqaction}
\end{align}
where $F'_{\mu\nu} \equiv \partial_\mu A'_\nu - \partial_\nu A'_\mu$ and $F_{\mu\nu} \equiv \partial_\mu A_\nu - \partial_\nu A_\mu$. 
The kinetic mixing between $A'_\mu$ and $A_\mu$ is characterized by the dimensionless parameter $\varepsilon$. 
The mass of the dark photon is denoted by $m_{A'}$. 
We work in natural units where the speed of light, the reduced Planck constant, and the vacuum permittivity are set to unity.

The mixing parameter $\varepsilon$ is assumed to be sufficiently small.
Then, the dark photon approximately satisfies the Proca equation $( \partial_\nu \partial^\nu - m_{A'}^2 ) A^{\prime \mu} = 0$ with $\partial_\mu A^{\prime \mu} = 0$.
Since the dark photon DM is non-relativistic, it can be described as a coherently oscillating wave with an angular frequency approximately equal to its mass $m_{A'}$.
The corresponding frequency $f_{A'}$ reads
\begin{align}
    {f}_{A'} &= \frac{m_{A'}}{2\pi} = 2.42 \, \text{Hz} \,  \bigg( \frac{m_{A'}}{10^{-14} \, \text{eV}} \bigg) \,.
    \label{eqfA'}
\end{align}
The coherence length is estimated as $\lambda_{\text{coh}} = 2\pi / (m_{A'} v) \sim 1 \times 10^{8} \, \text{km} \,(10^{-14} \, \text{eV}/ m_{A'}) $ using a typical velocity $v \sim 10^{-3}$. This length scale far exceeds the Earth's size for masses of interest.
Hence, within the coherence time $\tau_{\text{coh}} \sim 2\pi / (m_{A'} v^2) \sim 1.2 \times 10^{2} \, \text{hr} \, (10^{-14} \, \text{eV}/ m_{A'})$, the spatial components of the dark photon field at the Earth can be expressed as
\begin{align}
    \bm{A}'(t) = \hat{\bm{A}}' e^{-im_{A'} t} \,,
    \label{eqA'2}
\end{align}
where $\hat{\bm{A}}' = (\hat{A}'_j)$ $(j=x,y,z)$ is a constant vector.
(While the physical field is real, we express the solution in complex form for convenience.)

\section{Terrestrial magnetic field induced by dark photon DM}

From the action \eqref{eqaction}, the equation of motion for the ordinary photon $A_\mu$ is given by 
\begin{align}
    \partial_\nu F^{\nu \mu} = - \varepsilon \, \partial_\nu F^{\prime \nu \mu} \,. 
    \label{eqMax1}
\end{align}
This shows that, in the presence of the dark photon $A'_\mu$, an effective current $J_{\text{eff}}^\mu \equiv \varepsilon \, \partial_\nu F^{\prime \nu \mu}$ arises. 
Substituting \eqref{eqA'2} gives $\bm{J}_{\text{eff}} = \varepsilon \, m_{A'}^2 \hat{\bm{A}}' e^{-i m_{A'} t}$, which acts as a source for the observable electromagnetic field.
Let us estimate the order of magnitude of the induced magnetic field following the discussion in \cite{Fedderke:2021aqo}.
Suppose that the current $\bm{J}_{\text{eff}}$ penetrates an area of radius $R$.
Integrating the Amp\`{e}re--Maxwell law $\bm{\nabla} \times \bm{B} \sim \bm{J}_{\text{eff}}$ over this area yields $B R \sim \varepsilon m_{A'}^2 |\hat{\bm{A}}'| R^2 \sim \varepsilon m_{A'} \sqrt{\rho_{A'}} R^2$, where $\rho_{A'}$ denotes the energy density of the dark photon DM, and we use $\rho_{A'} \sim m_{A'}^2 |\hat{\bm{A}}'|^2 / 2$.
In our case, the radius $R$ corresponds to the Earth's radius $R_{\text{E}} = 6.4 \times 10^3 \, \text{km}$. Hence, the magnetic field induced on the Earth is roughly estimated as
\begin{align}
    B_{A'} &\sim \varepsilon \, m_{A'} R_{\text{E}} \sqrt{\rho_{A'}}
    \notag \\
    & \sim 3 \times 10^{-2} \, \text{pT}\bigg( \frac{\varepsilon}{10^{-8}} \bigg) \bigg( \frac{m_{A'}}{10^{-14}\, \text{eV}} \bigg) \sqrt{ \frac{\rho_{A'}}{0.3 \, \text{GeV}/\text{cm}^3} }\,,
    \label{estimate}
\end{align}
where we use the local DM density $\rho_{\text{DM}} = 0.3 \, \text{GeV}/\text{cm}^3$ \cite{Read:2014qva} as the fiducial value for $\rho_{A'}$.
Note that this magnetic field oscillates at the frequency of \eqref{eqfA'}. 
Such a signal would manifest as a sharp and persistent spectral line, distinguishable from transient noise.

We now proceed to a more accurate treatment.
A simplified model of the atmospheric environment would assume a vacuum with vanishing conductivity, and set a perfectly conducting upper boundary at a certain altitude to mimic the ionosphere.
However, such a model predicts a formal divergence in the amplitude of the induced electromagnetic field at the resonance frequencies of the Earth-ionosphere cavity (see Fig.~\ref{fig:B_amp_comparison} in Appendix).
Therefore, incorporating atmospheric conductivity is essential for accurately predicting the induced signal.
To include conductivity, we consider the following Maxwell equations with the effective current $\bm{J}_{\text{eff}}$,\footnote{The relative magnetic permeability is approximated as unity.}
\begin{align}
    \bm{\nabla} \cdot \bm{D} &= 0\,, 
    \label{eqMax11}\\
    \bm{\nabla} \times \bm{B} - \dot{\bm{D}} &= \varepsilon \, m_{A'}^2 \hat{\bm{A}}'  e^{-im_{A'} t} \, ,
    \label{eqMax12}
\end{align}
where an overdot denotes the time derivative.
Here, the displacement field $\bm{D}$ is related to the electric field $\bm{E}$ via the refractive index $n$ as $\bm{D} = n^2 \bm{E}$.
Since the source term in \eqref{eqMax12} oscillates as $e^{-im_{A'}t}$, we consider electromagnetic fields with angular frequency $m_{A'}$. 
The refractive index $n$ is then given by $n^2 = 1 + i (\sigma /  m_{A'})$,
where $\sigma$ denotes the conductivity.
We assume that the conductivity is a function of the radial coordinate $r$ measured from the Earth's center, $\sigma = \sigma (r)$.
In this work, we adopt the atmospheric conductivity profile from \cite{kudintseva2016ac}, which gradually increases from $10^{-14}\, \text{S/m}$ at the lowest atmosphere to $10^{-3}\, \text{S/m}$ at an altitude of $99\,\text{km}$.

The magnetic field $\bm{B}_{A'}$ induced in the atmosphere through \eqref{eqMax12} can be expressed as\footnote{
In addition to the frequency $f_{A'} = m_{A'}/(2\pi)$, the Earth's rotation generates sidebands shifted by $f_{\text{day}} = 1/\text{day} = 1.2 \times 10^{-5} \, \text{Hz}$ \cite{Fedderke:2021aqo}.
However, since the data segment length is fixed to 8 hours in our analysis, these sidebands fall within the frequency bin width and are thus unresolvable.
}
\begin{align}
    \bm{B}_{A'}(t,r,\Omega) =
    - i \, \varepsilon \, m_{A'}  b(x) \, e^{-im_{A'} t} \sum_{m=-1}^{+1} \hat{A}_m'\bm{\Phi}_{1,m}(\Omega) \,.
    \label{eqBA'1}
\end{align}
Here, $\Omega=(\theta, \phi)$ denotes the spherical coordinates, $b(x)$ is a function of the dimensionless variable $x \equiv m_{A'}r$, and $\hat{A}'_m$ are particular linear combinations of the dark photon components $\hat{A}'_j$ (see Appendix) normalized as $\sum_{m} \braket{|\hat{A}'_m|^2} = 2\rho_{A'}/m_{A'}^2$, where the angle bracket represents the ensemble average over various realizations of field amplitudes and phases \cite{Nakatsuka:2022gaf}. 
The $\bm{\Phi}_{\ell, m} \equiv \bm{r} \times \bm{\nabla} Y_{\ell, m}$ is one of the vector spherical harmonics with $Y_{\ell, m}$ being the scalar spherical harmonics \cite{barrera1985vector}.
Note that the induced magnetic field contains only the $\ell = 1$ mode due to the dipolar nature of the dark photon, and the magnetic field is perpendicular to the direction of the dark photon vector.
The function $b(x)$ is obtained as the solution to the following equation (see Appendix), 
\begin{align}
    \bigg( \frac{d}{dx} \frac{1}{n^2} \frac{d}{dx} + 1 - \frac{2}{n^2 x^2} \bigg) (x\,b) &= - i \sqrt{\frac{4\pi}{3}} \, x \, \frac{d}{dx}\frac{1}{n^2} \,,
    \label{eqindB2-3}
\end{align}
subject to the following boundary conditions.
Since the Earth's surface acts as a conducting boundary, the tangential component of $\bm{E}$ must vanish there.
At the lower ionosphere, around an altitude of $99\, \text{km}$, we impose an outgoing-wave boundary condition.
We also model the atmosphere as a stack of layers with constant conductivity, allowing the equation to be solved semi-analytically.
We follow the treatment in \cite{Taruya:2025bhe}, which analyzed the magnetic field induced by axion DM.

\begin{figure}[tb]
    \centering
    \includegraphics[width=\columnwidth]{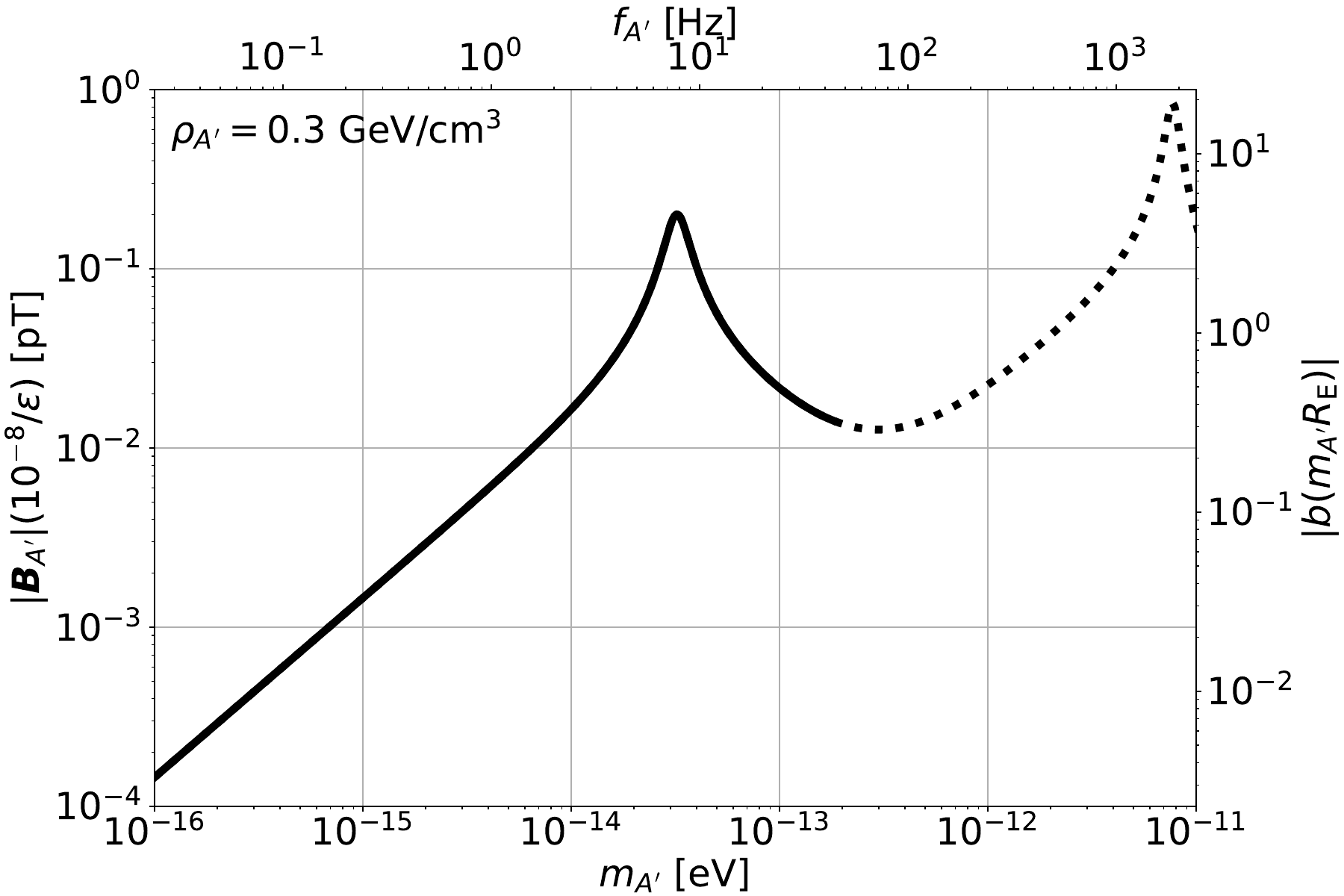}
    \caption{The expected amplitude of the magnetic field $|\bm{B}_{A'}|$ induced by dark photon DM (left vertical axis) and the corresponding value of $|b(m_{A'}R_{\text{E}})|$ (right vertical axis) are shown, assuming the DM energy density $\rho_{A'} = 0.3 \, \text{GeV}/\text{cm}^3$.
    The left axis is normalized by $\varepsilon/10^{-8}$, where $\varepsilon$ is the kinetic mixing parameter.
    The lower and upper horizontal axes are the dark photon mass $m_{A'}$ and the corresponding frequency $f_{A'} = m_{A'}/(2\pi)$, respectively.
    The dotted curve indicates the region $f_{A'} > 44\,\text{Hz}$, which is not included in data analysis of this work.
    }
    \label{fig:B_amp}
\end{figure}

The expected amplitude of the induced magnetic field at the Earth's surface ($r=R_{\text{E}}$) is defined by $| \bm{B}_{A'} | \equiv \sqrt{ \braket{ |\bm{B}_{A'}(t, R_{\text{E}}, \Omega) |^2 }}$ in terms of the ensemble average.
Assuming equipartition among the three spatial components of the dark photon field, we obtain
\begin{align}
    |\bm{B}_{A'}| &= \frac{\varepsilon \sqrt{\rho_{A'}}}{\sqrt{\pi}} \big|b(m_{A'} R_\text{E})\big|
    \notag \\
    &= 4.4 \times 10^{-2} \, \text{pT} 
    \cdot \big| b(m_{A'} R_\text{E}) \big|
    \notag \\
    &\quad \times \bigg( \frac{\varepsilon}{10^{-8}} \bigg)  \sqrt{ \frac{\rho_{A'}}{0.3 \, \text{GeV}/\text{cm}^3} }\,,
    \label{predB}
\end{align}
which is shown in Fig.~\ref{fig:B_amp} as a function of the dark photon mass $m_{A'}$.\footnote{In Fig.~\ref{fig:B_amp}, the atmospheric conductivity profile from \cite{kudintseva2016ac} is used.
We verified that using different conductivity profiles from \cite{galuk2018amplitude,nickolaenko2016vertical} alters the prediction by less than 10\% for $m_{A'} \lesssim 3 \times 10^{-12}\, \text{eV}$. See Appendix for details and for a discussion of the potential impact of geographic and long-term temporal variations in the conductivity.
}
For masses $m_{A'} \lesssim 10^{-14} \, \text{eV}$, we find $b(m_{A'}R_{\text{E}}) \approx i \sqrt{\pi/3} \cdot m_{A'} R_{\text{E}}$, which reproduces the mass dependence derived in \eqref{estimate}. In this regime, the behavior of the magnetic field amplitude differs from the axion DM case, where the amplitude becomes independent of the axion mass \cite{Arza:2021ekq, Taruya:2025zql, Taruya:2025bhe}.
For larger masses, a resonant amplification appears around $m_{A'} \approx 3.2 \times 10^{-14}\, \text{eV}$, corresponding to a wavelength $\lambda_{A'} \approx 3.8 \times 10^{4}\, \text{km}$, where $|b(m_{A'}R_{\text{E}})| \approx 4.6$.
This behavior arises from the confinement of electromagnetic waves with wavelengths close to Earth's circumference ($2\pi R_{\text{E}} = 4.0 \times 10^4 \,\text{km}$) within the effective cavity formed by the Earth's surface and the ionosphere, analogous to the Schumann resonance known for extremely low frequency radio waves \cite{Schumann1952a, Schumann1952b, sentman2017schumann}.
Another resonance peak is observed around $m_{A'} \approx 7.7 \times 10^{-12}\, \text{eV}$, corresponding to a wavelength $\lambda_{A'} \approx 1.6 \times 10^2 \, \text{km}$, which is comparable to the ionospheric altitude.
Note that, in the case of axion DM, the signal is significantly suppressed for masses $m_{\text{a}} \gtrsim 10^{-13} \, \text{eV}$ \cite{Taruya:2025zql, Taruya:2025bhe}, in contrast to the dark photon case.

\section{Data analysis and constraints}

Based on the theoretical prediction obtained above, we search for the dark photon DM signal using terrestrial magnetic field data.
The predicted signal has a monochromatic spectrum with a frequency determined by the dark photon mass, similar to the case of axion DM \cite{Taruya:2025zql, Taruya:2025bhe}. This similarity allows the dark photon DM search to be performed in parallel with the previous axion DM search \cite{Taruya:2025zql, Nishizawa:2025xka}.
Specifically, in \cite{Nishizawa:2025xka}, the axion-induced magnetic field was expressed as $\bm{B}(t) = \bm{R}(f_{\text{a}}) \, g_{\text{a}\gamma} \, e^{-2\pi i f_{\text{a}} t}$, where $g_{\text{a}\gamma}$ is the axion-photon coupling constant, and $2\pi f_{\text{a}} = m_{\text{a}}$ with the axion mass $m_{\text{a}}$, and $\bm{R}(f_{\text{a}})$ is a mass-dependent function obtained by solving the Maxwell equation in the presence of axions. 
For the dark photon search, we replace $f_{\text{a}} \to f_{A'}$, $g_{\text{a}\gamma} \to \varepsilon$, and $\bm{R}(f_{\text{a}}) \to \bm{R}_{A'}(f_{A'})$, where $\bm{R}_{A'}(f_{A'})$ satisfies $\braket{|\bm{R}_{A'}(f_{A'})|^2} = (\rho_{A'}/\pi) \, |b(m_{A'}R_{\text{E}})|^2$.
Note that $\bm{R}(f_{\text{a}})$ and $\bm{R}_{A'}(f_{A'})$ differ in their mass dependence.

The data analysis for the monochromatic signals follows the procedure in \cite{Nishizawa:2025xka} and uses the same data product.
Below, we summarize the essential steps.
We use data from the Eskdalemuir Observatory, recorded between September 1, 2012, and November 4, 2022 \cite{BritishGeologicalSurvey, beggan2018observation}. 
The dataset consists of two channels oriented in different horizontal directions with a sampling rate of $100\,\text{Hz}$.
From the full dataset, we select data portions recorded continuously for over one month, and divide them into segments of duration $T_{\text{seg}} = 8\,\text{hr}$ each.
This yields a total of $N_{\text{seg}} = 9909$ segments.
For each data segment, the power spectral density (PSD) is computed from both channels.
The frequency bin width is $\Delta f \approx T_{\text{seg}}^{-1} \approx 3.5 \times 10^{-5}\, \text{Hz}$, ensuring that the dark photon signal with a frequency width $\Delta f_{A'} \sim v^2 \cdot f_{A'}$ remains within a single bin for $f_{A'} \lesssim 35\, \text{Hz}$.
The PSDs are then stacked with weights determined by the noise variance estimated from surrounding frequency bins.
By subtracting a smooth component from the sum, we obtain a weighted-average differential PSD, $\hat{s}(f)$, shown in Fig.~3 of \cite{Nishizawa:2025xka}.
The resulting PSD shows many spectral lines at integer frequencies, which are likely artificial. We therefore exclude $\pm 1000$ bins around each integer frequency.

To search for the DM-induced signal, we evaluate the signal-to-noise ratio (SNR) for each frequency bin. 
The noise variance is estimated from bins surrounding each target frequency.
The SNR is computed from $0.01 \, \text{Hz}$ to $44 \, \text{Hz}$, where the upper limit is set by the Nyquist frequency.
We identify signal candidates by the following three criteria: 1) the SNR exceeds 2 in all-year data; 2) the SNR in each year exceeds a threshold weighted by the inverse noise variance for that year; and 3) the frequency differs from any multiple of $0.05\,\text{Hz}$ by more than $10^{-3}\,\text{Hz}$.
Criterion 2) ensures the persistency of the signal, and criterion 3) eliminates spectral lines that are likely produced by analog filters or the digitizer.

Applying the above criteria, we identify 342 candidate frequencies.
See also \cite{Nishizawa:2025xka} for candidates obtained using different SNR thresholds. There are 65 candidates with $\text{SNR}>3$ and 31 candidates with $\text{SNR}>5$ in all-year data. 
Note that these appear as prominent spectral lines, but our analysis cannot distinguish whether they originate from dark photon DM or axion DM.
The previous work \cite{Friel:2024shg}, using SuperMAG 1-sec sampling data, reported five candidate frequencies below $1\,\text{Hz}$, possibly from dark photon DM.
In the same frequency range, our analysis identified five candidates with $\text{SNR}>2$ (see Table IV in \cite{Nishizawa:2025xka}); however, none of them coincides with their candidates. 
Although our sensitivity is slightly lower than theirs at $\lesssim 0.3\, \text{Hz}$, we find no significant signals and place constraints at four of their five candidate frequencies. See Appendix for details.

Apart from the above candidates, no significant signals are found. From this, we obtain constraints on the mixing parameter $\varepsilon$, for a given dark photon DM density $\rho_{A'}$.
The 95\% C.L.~upper limit on $\varepsilon$ is derived from
\begin{align}
    \int_{-\infty}^{\hat{s}_{\text{obs}}(f_{A'})} d\hat{s} \, p[ \hat{s}(f_{A'}) | \varepsilon ] = 0.05 \,. 
    \label{upperlimit}
\end{align}
Here, $\hat{s}_{\text{obs}}$ is the observed weighted-average differential PSD, which is converted to $\varepsilon$ by equating it with $\braket{ \hat{s}(f_{A'}) } \equiv 2T_{\text{seg}} \, \varepsilon^2 \braket{|\bm{R}_{A'}(f_{A'})|^2}$.
The $p[ \hat{s}(f_{A'}) | \varepsilon ]$ is the probability density distribution of the weighted-average differential PSD in the presence of dark photon DM, constructed from the data in the surrounding frequencies and the signal prediction.
The resulting constraints on $\varepsilon$ with $\rho_{A'} = 0.3 \, \text{GeV/cm}^3$ are shown as the blue region in Fig.~\ref{fig:darkphoton_constraint}.
The limit is as strong as $\varepsilon \lesssim 5 \times 10^{-11}$ around $m_{A'} \approx 3 \times 10^{-14}\, \text{eV}$, corresponding to the resonance peak of the induced magnetic field shown in Fig.~\ref{fig:B_amp}.

\section{Conclusion and discussion}

In this paper, we present a novel prediction for the terrestrial magnetic field induced by dark photon DM, incorporating the effect of atmospheric conductivity and capturing the resonance near $8 \, \text{Hz}$ within the Earth-ionosphere cavity. 
This enables a search for dark photon DM with masses $m_{A'} \lesssim 10^{-13}\,\text{eV}$ using magnetic field data below $100\,\text{Hz}$, offering a new and complementary probe to existing approaches \cite{Sulai:2023zqw, Jiang:2023jhl, Bloch:2023wfz}.
Assuming the fiducial DM energy density, we derive stringent upper limits on the kinetic mixing parameter $\varepsilon$, surpassing those from existing ground-based direct experiments in the mass range of $1 \times 10^{-15} \, \text{eV} \lesssim m_{A'} \lesssim 2 \times 10^{-13} \, \text{eV}$, as shown in Fig.~\ref{fig:darkphoton_constraint}.
While strong indirect constraints have been inferred from cosmological probes for $m_{A'} \gtrsim 10^{-14}\, \text{eV}$ \cite{Caputo:2020bdy, Witte:2020rvb}, our results provide direct and independent bounds.
Notably, in the mass range of $1 \times 10^{-15}\, \text{eV} \lesssim m_{A'} \lesssim 8 \times 10^{-15} \, \text{eV}$, our limit is the most stringent to date.

In addition to setting constraints, we found a number of signal candidates appearing as prominent spectral lines.
However, our single-site PSD analysis does not allow us to determine whether they originate from dark photon DM or axion DM.
To identify the origin, it would be necessary to examine the global magnetic field profile, as axion-induced fields follow the geomagnetic field, whereas dark photon-induced fields do not \cite{Fedderke:2021rrm, Arza:2021ekq}.

We also note the potential impact of the stochastic nature of DM on the derived parameter constraints \cite{Centers:2019dyn,Lisanti:2021vij}. As demonstrated in \cite{Nakatsuka:2022gaf}, this effect becomes negligible when the observation time exceeds roughly ten times the coherence time. In our case, the constraints remain robust for DM masses above $2 \times 10^{-15} \,\text{eV}$ \cite{Nishizawa:2025xka}.

Finally, we note the potential utility of our magnetic field prediction in the higher frequency regime $f_{A'} \gtrsim 100 \, \text{Hz}$, which was not analyzed in this work. 
In particular, the dark photon-induced magnetic field exhibits an additional resonance near $f_{A'} \approx 2 \, \text{kHz}$ (see Fig.~\ref{fig:B_amp}), which is less pronounced in the axion-induced case. Thus, magnetic field measurements in the kHz range could serve as an effective probe of dark photon DM within the corresponding mass window.

\section*{Acknowledgments}
We are grateful to C. Beggan for  kindly providing the data at the Eskdalemuir observatory and for his valuable comments. 
This work was supported in part by JSPS KAKENHI Grant Numbers
JP24KJ0117, JP25K17389 (KN), 
JP23K03408, JP23H00110, JP23H04893 (AN),
JP20H05861, JP23K20844, JP23K25868 (AT), 
and JP21K03580, JP25K07288 (YH).

\appendix


\section{Derivation of master equation}

We present the derivation of the master equation describing electromagnetic fields induced by dark photon DM in the Earth's atmosphere.
We extend the analysis of \cite{Fedderke:2021aqo} to the case of radially varying conductivity.
Taking the time derivative of \eqref{eqMax12} and using $\bm{D} = n^2 \bm{E}$ and the Bianchi identity $\dot{\bm{B}} + \bm{\nabla} \times \bm{E} = 0$, we obtain the equation for the electric field $\bm{E}$ as
\begin{align}
    - \Delta \bm{E} - n^2 m_{A'}^2 \bm{E} + \bm{\nabla} (\bm{\nabla} \cdot \bm{E}) 
    = i \, \varepsilon \, m_{A'}^3 \hat{\bm{A}}' e^{-im_{A'} t} \,,
    \label{eqMax21}
\end{align}
where $\Delta \equiv \bm{\nabla}^2$ is the Laplacian, and the time dependence of $\bm{E} \propto e^{-im_{A'} t}$ is considered.
To solve the above equation based on the spherical symmetry of the Earth, we introduce the vector spherical harmonics as $\bm{Y}_{\ell m} \equiv \hat{\bm{r}} \, Y_{\ell m}$, $\bm{\Psi}_{\ell m} \equiv r \bm{\nabla} Y_{\ell m}$, and $\bm{\Phi}_{\ell m} \equiv \bm{r} \times \bm{\nabla} Y_{\ell m} $ \cite{barrera1985vector}.
Here, $Y_{\ell m}(\theta, \phi)$ represents the ordinary (scalar) spherical harmonics, $\bm{r} \equiv r (\sin \theta \cos \phi, \sin \theta \sin \phi, \cos \theta)$ is the radial vector, and $\hat{\bm{r}} \equiv \bm{r}/ r$ is the radial unit vector.

The displacement vector $\bm{D}$ compatible with the transverse condition \eqref{eqMax11} can be decomposed into two modes, the TE (transverse-electric) mode and the TM (transverse-magnetic) mode, $\bm{D} = \bm{D}_{\text{TE}} + \bm{D}_{\text{TM}}$.
These modes are expressed in terms of the vector spherical harmonics as 
\begin{align}
    \bm{D}_{\text{TE}} &= \sum_{\ell, m} f_{\ell m}(r) \, \bm{\Phi}_{\ell m} \, e^{-i m_{A'} t} \,,
    \label{eqD2}\\
    \bm{D}_{\text{TM}} &= \sum_{\ell, m} \frac{1}{m_{A'}} 
    \bm{\nabla} \times (g_{\ell m}(r) \, \bm{\Phi}_{\ell m}) \, e^{-im_{A'} t} 
    \notag \\
    &= \sum_{\ell, m} \frac{1}{m_{A'}} 
    \bigg[ - \frac{\ell (\ell + 1)}{r} g_{\ell m}(r) \, \bm{Y}_{\ell m}
    \notag \\
    &\qquad
    - \frac{1}{r} \frac{d}{dr} (r g_{\ell m}(r)) \, \bm{\Psi}_{\ell m} \bigg] e^{-i m_{A'} t}  \,,
    \label{eqD3}
\end{align}
where $f_{\ell m}(r)$ and $g_{\ell m}(r)$ are radial functions.
The electric field $\bm{E}$ is correspondingly decomposed as $\bm{E} = \bm{E}_{\text{TE}} + \bm{E}_{\text{TM}}$, where $\bm{E}_{\text{TE}} = \bm{D}_{\text{TE}} / n^2$ and $\bm{E}_{\text{TM}} = \bm{D}_{\text{TM}} / n^2$.

We can also express the right-hand side of \eqref{eqMax21} using the vector spherical harmonics \cite{Fedderke:2021aqo}.
Using the explicit forms of several vector spherical harmonics listed in Appendix D of \cite{Fedderke:2021aqo}, we find that the unit vectors along the Cartesian coordinate axes $\{ \hat{\bm{x}}, \hat{\bm{y}}, \hat{\bm{z}} \}$ are expressed as
\begin{align}
    \hat{\bm{x}} &= - \sqrt{ \frac{2\pi}{3} } ( \bm{Y}_{11} - \bm{Y}_{1,-1} + \bm{\Psi}_{11} - \bm{\Psi}_{1,-1} ) \,, 
    \label{eqbx2}\\
    \hat{\bm{y}} &= i \sqrt{ \frac{2\pi}{3} } ( \bm{Y}_{11} + \bm{Y}_{1,-1} + \bm{\Psi}_{11} + \bm{\Psi}_{1,-1} ) \,,  
    \label{eqby2}\\
    \hat{\bm{z}} &= \sqrt{ \frac{4\pi}{3} } ( \bm{Y}_{10} + \bm{\Psi}_{10} ) \,. 
    \label{eqbz2}
\end{align}
Therefore, the constant vector $\hat{\bm{A}}'$ has the following expression, 
\begin{align}
    \hat{\bm{A}}'
    &= \hat{A}_x' \hat{\bm{x}} + \hat{A}_y' \hat{\bm{y}} + \hat{A}_z' \hat{\bm{z}} 
    = \sum_{m=-1}^{+1} 
     \hat{A}_m' (\bm{Y}_{1m} + \bm{\Psi}_{1m}) \,,
     \label{eqA'15}
\end{align}
where $\hat{A}'_{+1}$, $\hat{A}'_0$, and $\hat{A}'_{-1}$ are defined in terms of the Cartesian components $\hat{A}'_j = (\hat{A}'_x, \hat{A}'_y, \hat{A}'_z)$ as
\begin{align}
    \hat{A}_{+1}' &\equiv - \frac{1}{\sqrt{2}} (\hat{A}_x' - i \hat{A}_y') \,, 
    \label{eqA'12}\\
    \hat{A}_0' &\equiv \hat{A}_z' \,, 
    \label{eqA'13}\\
    \hat{A}_{-1}' &\equiv \frac{1}{\sqrt{2}} (\hat{A}_x' + i \hat{A}_y') \,.
    \label{eqA'14}
\end{align}

From \eqref{eqA'15}, the right-hand side of \eqref{eqMax21} has only components with $\bm{Y}_{1m}$ and $\bm{\Psi}_{1m}$.
Therefore, the electric field of the TE mode, which is the linear combination of $\bm{\Phi}_{\ell m}$, cannot be induced by dark photon DM. 
For this reason, we focus only on the TM mode, and we write the induced electric field $\bm{E}_{\text{TM}}$ and magnetic field $\bm{B}_{\text{TM}}$ as $\bm{E}_{A'}$ and $\bm{B}_{A'}$, respectively.

Substituting \eqref{eqD3} into \eqref{eqMax21}, we obtain the equations for the radial functions $g_{\ell m}(r)$. 
Since the dark photon DM contributes only to the $\ell = 1$ modes as shown in \eqref{eqA'15}, it suffices to study $g_{1 m}(r)$.
Extracting the coefficient of $\bm{Y}_{1m}$, we obtain the following equation for $g_{1 m}(r)$,
\begin{align}
    &\frac{d^2 g_{1m}}{dr^2}  + \bigg[ n^2 \bigg( \frac{d}{dr} \frac{1}{n^2} \bigg)  + \frac{2}{r} \bigg] \frac{d g_{1m}}{dr}
    \notag \\
    &+ \bigg[  \frac{n^2}{r}  \bigg( \frac{d}{dr} \frac{1}{n^2} \bigg) + n^2 m_{A'}^2  - \frac{2}{r^2} \bigg]  g_{1m}
    = i \, \varepsilon \,m_{A'}^4 \sqrt{\frac{\pi}{3}} \, r n^2 \hat{A}_m' \,,
    \label{eqg1m1}
\end{align}
where the refractive index $n$ is assumed to be a function of the radius, $n=n(r)$.
We also see that the equation obtained by extracting the coefficient of $\bm{\Psi}_{1m}$ is automatically satisfied under \eqref{eqg1m1}.

Noting that $g_{1m}$ is sourced from the right-hand side of \eqref{eqg1m1} proportional to $\varepsilon \hat{A}'_m$, it is legitimate to factorize $\varepsilon \hat{A}'_m$ from $g_{1m}$.
Furthermore, to simplify the equation, we introduce a new function $u$ by
\begin{align}
    g_{1m} \equiv \varepsilon \hat{A}'_m \, \frac{u}{r}\,.
    \label{eqg0}
\end{align} 
Then, equation \eqref{eqg1m1} is rewritten as 
\begin{align}
    \frac{d}{dx} \bigg( \frac{1}{n^2} \frac{d u}{dx} \bigg) + \bigg( 1 - \frac{2}{n^2 x^2} \bigg) u &= i \, \sqrt{\frac{\pi}{3}} \,x^2 \,,
    \label{equ1m2}
\end{align}
which is the master equation to be solved.

Given the electric field $\bm{E}_{A'}$, the corresponding magnetic field can be provided by the Bianchi identity as 
\begin{align}
    \bm{B}_{A'} = \frac{1}{i\, m_{A'}} \bm{\nabla} \times \bm{E}_{A'} \,,
    \label{eqindB1}
\end{align}
where the time-dependence of $\bm{B}_{A'} \propto e^{-im_{A'} t}$ is considered.
Substituting the $\ell = 1$ part of $\bm{E}_{A'}$ into the right-hand side and using \eqref{eqg1m1}, the induced magnetic field is expressed as 
\begin{align}
    \bm{B}_{A'} =
    - i \, \varepsilon \, m_{A'}  b(x) \, e^{-im_{A'} t} \sum_{m=-1}^{+1} \, \hat{A}_m'\bm{\Phi}_{1 m} \,,
    \label{eqindB3}
\end{align}
where we defined the function $b(x)$ with $x = m_{A'}r$ as
\begin{align}
    b(x) \equiv \frac{u(x)}{x} - i \, \sqrt{\frac{\pi}{3}} \, x \,.
    \label{eqindB2-2}
\end{align}
 From \eqref{equ1m2}, the function $b(x)$ satisfies 
\begin{align}
    \frac{d}{dx} \bigg( \frac{1}{n^2} \frac{d}{dx} (x \,b) \bigg) + \bigg( 1 - \frac{2}{n^2 x^2} \bigg) (x\,b) &= - \,i \,2 \sqrt{\frac{\pi}{3}} \, x \, \frac{d}{dx}\frac{1}{n^2} \,,
    \label{eqindB2-3-sup}
\end{align}
which appears in \eqref{eqindB2-3} of the main text.

\section{Different conductivity profiles}

\begin{figure}[tb]
    \centering
    \includegraphics[width=\columnwidth]{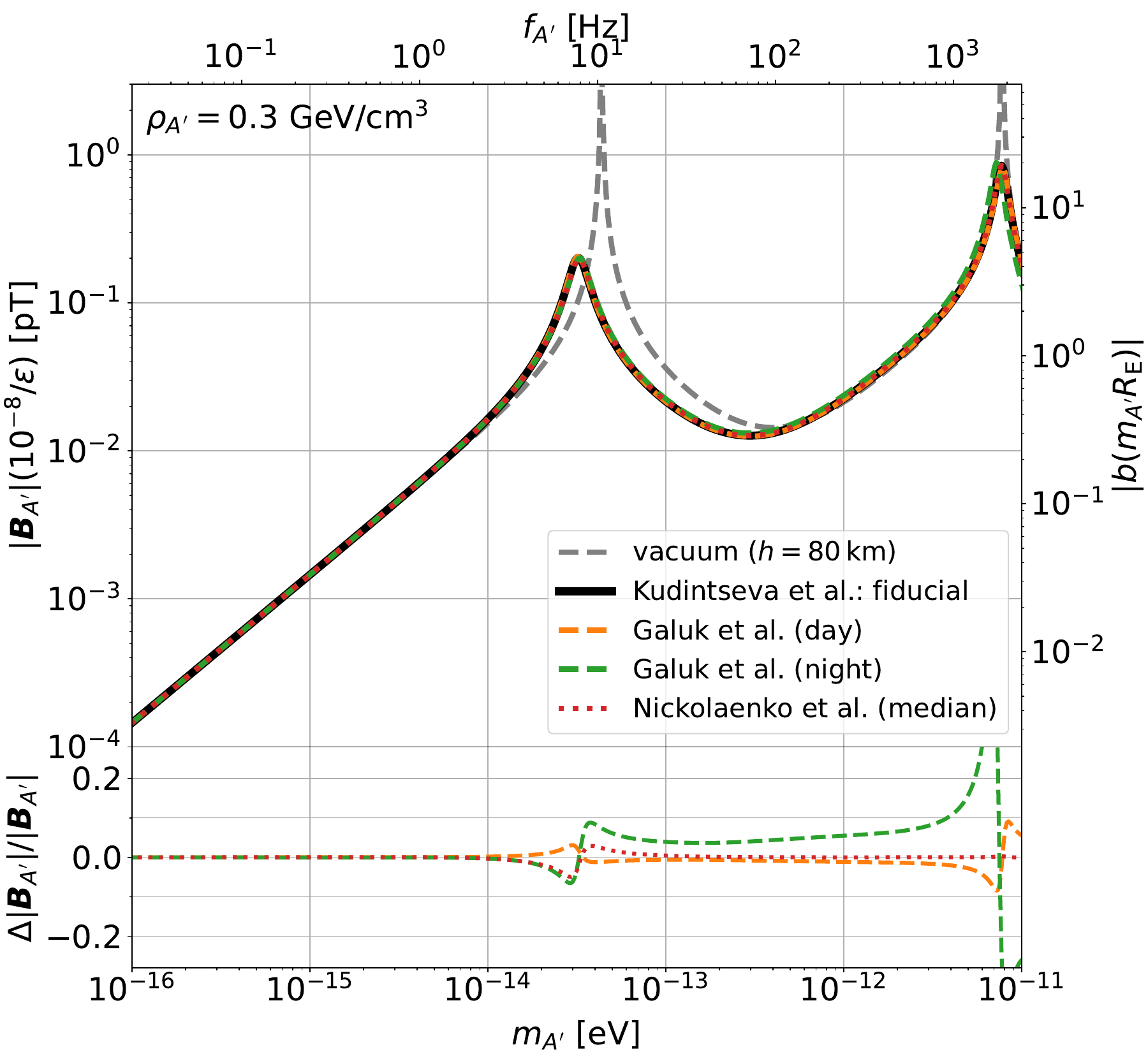}
    \caption{Top: Expected amplitudes of the induced magnetic field $|\bm{B}_{A'}|$ (left vertical axis) and the values of $|b(m_{A'}R_{\text{E}})|$ (right vertical axis) calculated using different radial profiles of atmospheric conductivity are shown.
    The black curve corresponds to the conductivity profile in \cite{kudintseva2016ac}.
    The yellow and green dashed curves use the dayside and nightside profiles from \cite{galuk2018amplitude}, respectively, while the red dotted curve corresponds to the median profile from \cite{nickolaenko2016vertical}.
    The gray dashed curve represents the result assuming a vacuum environment with a perfectly conducting upper boundary at an altitude of $h = 80 \, \text{km}$.
    All results assume $\rho_{A'}=0.3 \, \text{GeV/cm}^3$.
    Bottom: Deviations of the expected amplitude using each conductivity profile relative to that obtained with the profile in \cite{kudintseva2016ac} are shown.
    }
    \label{fig:B_amp_comparison}
\end{figure}

In our work, we use the radial profile of atmospheric conductivity up to an altitude of $99\,\text{km}$ from \cite{kudintseva2016ac} as the fiducial model to predict the magnetic field induced by dark photon DM.
To test the robustness of this prediction, we repeat the calculation using alternative conductivity profiles: the dayside and nightside profiles up to $110\,\text{km}$ from \cite{galuk2018amplitude}, and the median profile up to $98\,\text{km}$ from \cite{nickolaenko2016vertical}.
These profiles are illustrated in Fig.~2 of \cite{Taruya:2025bhe}.
The top panel of Fig.~\ref{fig:B_amp_comparison} shows the expected amplitude of the induced magnetic field, $|\bm{B}_{A'}|$, calculated using each conductivity profile.
The bottom panel of Fig.~\ref{fig:B_amp_comparison} shows the deviation of each result relative to that obtained using the profile in \cite{kudintseva2016ac}.
We find that the relative deviation remains below 10\% for $m_{A'} \lesssim 3 \times 10^{-12}\, \text{eV}$.

In the top panel of Fig.~\ref{fig:B_amp_comparison}, we also plot the magnetic field amplitude calculated while neglecting atmospheric conductivity, i.e., assuming a vacuum environment, as the gray dashed curve. 
In this calculation, the upper boundary is set at an altitude of $80\,\text{km}$ and assumed to be perfectly conducting.
Clearly, the amplitude diverges at the resonance points.
This demonstrates that incorporating finite atmospheric conductivity is essential for accurately predicting the induced magnetic field.
Note that, in this vacuum model bounded by perfectly conducting boundaries, the leading-order signal is independent of the boundary height $h$ in the limit $h \ll R_\mathrm{E}$, where $R_\mathrm{E}$ is the Earth's radius \cite{Fedderke:2021aqo,Arza:2021ekq,Taruya:2025bhe}. The dependence on $h$ appears as a subleading correction, suppressed by the factor $h / R_\mathrm{E}$, and becomes relevant only when the signal wavelength is comparable to or shorter than $h$. It is therefore negligible in the mass range considered in this work.

We comment on the potential impact of geographic and long-term temporal variations in atmospheric conductivity.
For the prediction of the magnetic-field signal induced at the Earth's surface, only the conductivity in the 60--100 km altitude range is relevant \cite{Taruya:2025bhe, bookNickolaenko}.
In this altitude range, the conductivity can vary by about 1--2 orders of magnitude across geographic locations \cite{kamiyama1957distribution, sheng2017dependence}.
As for temporal variations, at altitudes of order 100 km, the conductivity can undergo severalfold variations over decadal timescales associated with solar activity \cite{moen1993solar}. However, Fig.~\ref{fig:B_amp_comparison} demonstrates the robustness of the signal prediction against different conductivity profiles. In particular, although the day-night conductivity profiles in \cite{galuk2018amplitude} differ by 1--2 orders of magnitude in the 60--100 km altitude range, the predicted signal varies by less than 10\% for masses below $3 \times 10^{-12}$ eV. Therefore, within the mass range analyzed in this paper, geographic
and long-term temporal variations in conductivity are expected to affect the signal only at a comparable level. 
In contrast, for larger masses, for which the signal wavelength becomes comparable to or shorter than the ionospheric height (corresponding to the second peak in the magnetic-field signal), geographic variations in conductivity are expected to become important, and a more careful treatment would be necessary.

\section{Candidates in the previous work}
\label{app:candidates_comparison}

The previous work \cite{Friel:2024shg} using SuperMAG 1-sec sampling data reported five signal candidate frequencies below $1\,\text{Hz}$, possibly arising from dark photon DM. They are (i) $0.1100\, \text{Hz}$, (ii) $0.3053 \, \text{Hz}$, (iii) $0.6217\, \text{Hz}$, (iv) $0.9511\,\text{Hz}$, and (v) $0.9703\, \text{Hz}$.
Here, we describe the implications of our analysis for these candidates.
First, at frequencies (i) and (ii), we find no significant signals, although our sensitivity is slightly lower than that of \cite{Friel:2024shg}. 
Accordingly, we place 95\% C.L. upper limits on the kinetic mixing parameter as $\varepsilon < 6.5 \times 10^{-7}$ at (i), and $\varepsilon < 7.3 \times 10^{-8}$ at (ii), assuming $\rho_{A'} = 0.3 \, \text{GeV/cm}^3$.
Also, at frequencies (iii) and (iv), no significant signals are found.
Furthermore, our sensitivity at these frequencies is higher than or comparable to that of \cite{Friel:2024shg}. 
Therefore, the candidates (iii) and (iv) can be ruled out by our analysis.
Finally, the candidate frequency (v) lies outside our analysis window, as we eliminate frequencies within $\pm 0.0347 \, \text{Hz}$ of integer values.

\bibliography{refs_DP}

\end{document}